\documentclass[twocolumn,twocolappendix]{aastex631}

\usepackage{amssymb}
\usepackage{amsmath}
\usepackage[]{graphicx}
\usepackage{enumerate}
\usepackage{subeqnarray}
\usepackage{cases}
\usepackage{mathrsfs,amssymb}
\shorttitle{Self-modulation of CR electrons}
\shortauthors{Dogiel et al.}

\tightenlines

\begin{document}

\title{Self-modulation of cosmic rays in molecular clouds: Imprints in the radio observations}

\author{V.~A.~Dogiel}\affiliation{I.~E.~Tamm Theoretical Physics Division of P.~N.~Lebedev
Institute of Physics, 119991 Moscow, Russia}\author{D.~O.~Chernyshov}\affiliation{I.~E.~Tamm Theoretical Physics Division of
P.~N.~Lebedev Institute of Physics, 119991 Moscow, Russia}\author{A.~V.~Ivlev} \affiliation{Max-Planck-Institut f\"ur
extraterrestrische Physik, 85748 Garching, Germany}
\author{A.~M.~Kiselev}\affiliation{I.~E.~Tamm Theoretical Physics Division of P.~N.~Lebedev
Institute of Physics, 119991 Moscow, Russia}\author{A.~V.~Kopyev}\affiliation{I.~E.~Tamm Theoretical Physics Division of
P.~N.~Lebedev Institute of Physics, 119991 Moscow, Russia}

\correspondingauthor{A.~V.~Ivlev} \email{ivlev@mpe.mpg.de}

\begin{abstract}
We analyze properties of non-thermal radio emission from the Central Molecular Zone (CMZ) and individual molecular clouds,
and argue that the observed features can be interpreted in the framework of our recent theory of self-modulation of cosmic
rays (CRs) penetrating dense molecular regions. For clouds with gas column densities of $\sim10^{23}$~cm$^{-2}$, the theory
predicts depletion of sub-GeV CR electrons, occurring due to self-modulation of CR protons and leading to harder synchrotron
spectra in the sub-GHz range. The predicted imprints of electron depletion in the synchrotron spectra agree well with the
spectral hardening seen in available radio observations of the CMZ. A similar, but even stronger effect on the synchrotron
emission is predicted for individual (denser) CMZ clouds, such as the Sgr B2. However, the emission at frequencies above
$\sim$~GHz, where observational data are available, is completely dominated by the thermal component, and therefore new
observations at lower frequencies are needed to verify the predictions.
\end{abstract}

\keywords{cosmic rays -- turbulence -- radio continuum: ISM -- radiation mechanisms: non-thermal -- Galaxy: center}

%\date{\today}

%\maketitle

\section{Introduction}
\label{intro}

\citet{ivlev} and \citet{dog3} developed a self-consistent model of cosmic ray (CR) interactions with molecular clouds,
where CR penetration into the clouds is governed by the self-generated MHD turbulence. The resulting modulation of CR flux
occurs in the diffuse envelopes around the clouds, before CRs penetrate into denser regions. Inside the dense regions
MHD-fluctuations are damped by strong ion-neutral friction and particle propagate ballistically along the local magnetic
field lines.

Both the CR spectrum and spectrum of the turbulence obey the excitation-damping balance, where the growth rate of magnetic
fluctuations induced by CR flux in the envelope is compensated by damping of MHD waves due ion-neutral collisions
\citep{ivlev}. As a result, the CR spectrum inside the cloud develops a break at a certain energy $E_{\rm ex}(N_{{\rm
H}_2})$, which is an increasing function of the cloud column density $N_{{\rm H}_2}$ (and whose value is determined by the
shape of the interstellar spectrum). While the CR spectrum above $E_{\rm ex}$ remains unchanged, at lower energies the
self-modulation leads to significant depletion of CR density, and can generate a universal (i.e., independent of the
interstellar spectrum) flux of CRs entering the clouds.

The CR depletion is expected to affect the gamma-ray emission due to proton-proton collisions, produced in dense molecular
clouds of Central Molecular Zone (CMZ) as well as in molecular clouds in the vicinity of the Solar System
\citep[see][]{de16,dog3,tib21}. Unfortunately, the value of $E_{\rm ex}$ turns out to be relatively small for the parameters
of such clouds, typically varying between $0.3-3$~GeV \citep{dog3}. As a result, the gamma-ray emission may only be affected
at sub-GeV energies, where the resolution of Fermi-LAT becomes poor \citep{dog3}.

The MHD disturbances produced by dominant CR species, protons and nuclei, naturally also affect penetrating CR electrons.
The main difference between elections and hadrons of given energy is that electrons are able to produce gamma-ray emission
at a higher energy, since bremsstrahlung photons take away a significant fraction of electron's energy. Therefore, if the
electron bremsstrahlung was the dominant source of gamma-ray emission from a cloud, the effect of CR depletion would be seen
at higher emission energies and, thus, would be likely detectable. Indeed, \citet{yus13} pointed out that diffuse gamma-ray
emission from the Galactic center at $\sim$~GeV energies may be due to electron bremsstrahlung, as derived from a measured
non-thermal spectrum of radio emission \citep[see Figure~12 of][]{yus13}. On the other hand, it is generally believed that
gamma-ray emission at such energies must be dominated by proton-proton collisions \citep[see, e.g.,][and references
therein]{owen21}.

This suggests that it may be difficult to find the effect of CR depletion in the gamma-ray range, and therefore it is worth
shifting to lower emission energies. Indeed, features of the synchrotron emission from CR electrons modulated at $\sim$~GeV
energies should be observable at radio frequencies in $\sim$~GHz range. Due to a very high resolution of radio telescopes,
it is possible to observe individual clouds and, thus, to disentangle cloud emission from the background. The aim of this
paper is to describe and quantify fingerprints of CR self-modulation that are expected in radio observations of molecular
gas.

The paper is organized as follows. In Section~\ref{observ} we discuss available radio observations from dense molecular
clouds, in Section~\ref{spectrum} we present a concise summary of the CR self-modulation theory and derive the expression
for a modulated spectrum of CR electrons inside the clouds, in Section~\ref{expected} we obtain the expected synchrotron
emission from the CMZ and compare the theoretical results with observations, and in Section~\ref{conc} we summarize our
conclusions.

\section{Observed radio emission from the CMZ region and from individual molecular clouds}
\label{observ}

One of the most promising targets for studies of the CR modulation is the CMZ -- a region located near the Galactic center,
with dimensions about 500~pc~$\times$~200~pc in the Galactic plane and 30~pc in height \citep{ferriere07}. The gas is mostly
concentrated in cold molecular clouds \citep[see, e.g.,][]{mezger,ferriere07, ao16,mills18} with the volume filling factor
of $\sim10\%$ and the average volume density of $\sim10^4$~cm$^{-3}$ \citep{mills17,mills18}. The diffuse medium surrounding
the clouds has an average gas density of $\la 50$~cm$^{-3}$ \citep{oka05,oka19,riquel}. Thus, the average column density of
the CMZ, $N_{{\rm H}_2}^{\rm CMZ}$, is dominated by the contribution of dense clouds. Its value can be estimated as a
product of the gas density averaged over the CMZ volume, $\sim10^3$~cm$^{-3}$, multiplied with the CMZ height, which yields
$N_{{\rm H}_2}^{\rm CMZ}\sim 10^{23}$~cm$^{-2}$. The mean magnetic field strength in the Galactic center is estimated as
$B\sim 0.1$~mG \citep{crock10}, while in the clouds it can reach $B\sim 1$~mG \citep{ferriere09}.

\begin{figure}[h]
\centering
\includegraphics[width=.95\columnwidth]{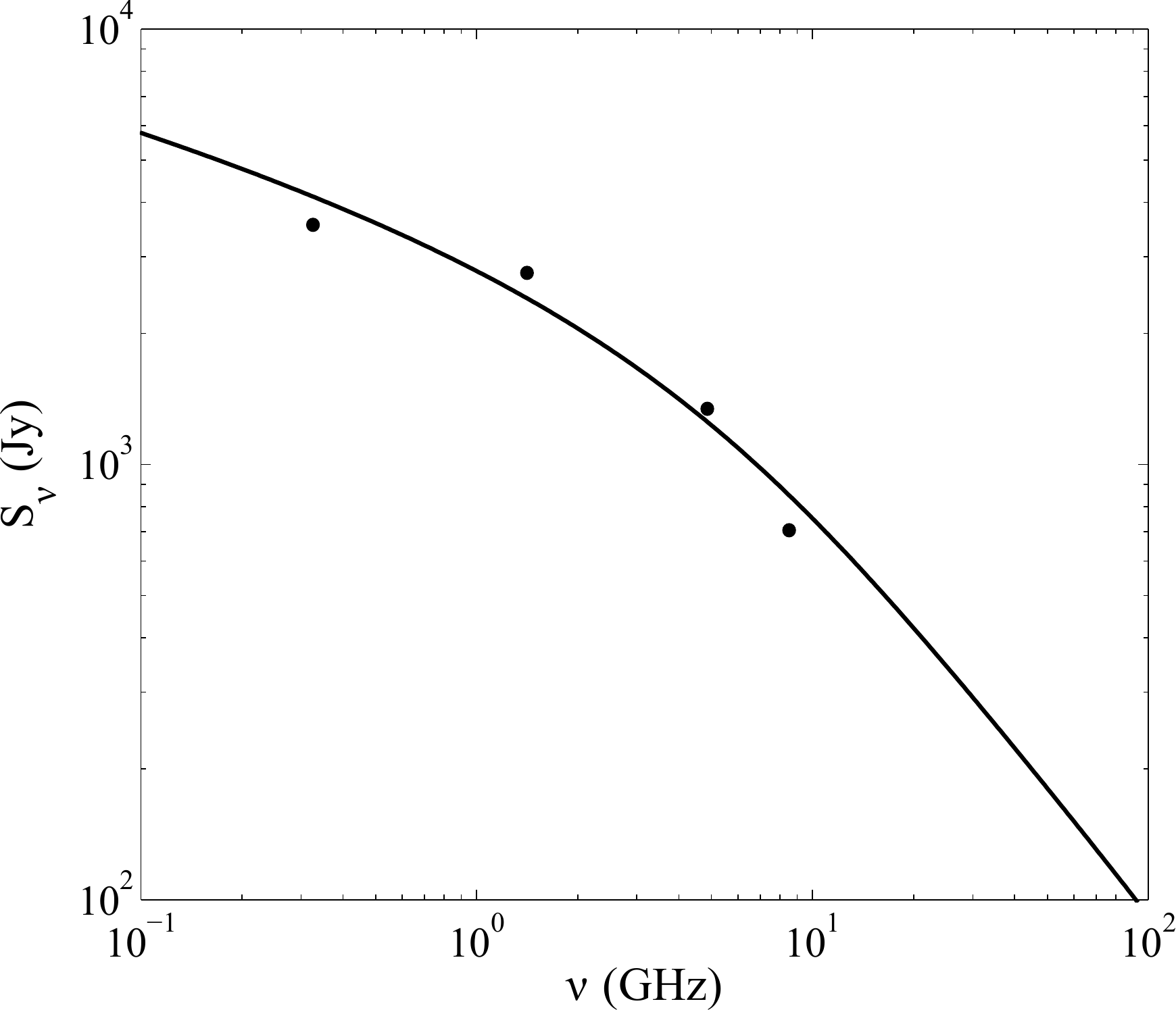}
\caption{Diffuse non-thermal radio emission measured by \citet{yus13} from the inner $2^\circ\times1^\circ$ region around the
Galactic center (bullets, statistical error bars are smaller than the symbols). The solid line shows the spectrum of
synchrotron emission predicted by our model (see Section~\ref{expected}).}
\label{fig1}
\end{figure}

\citet{yus13} found a strong spatial correlation between distributions of nonthermal radio sources and of molecular clouds
in the CMZ, and concluded that the nonthermal radio continuum is produced by relativistic electrons. It was estimated that
contribution of the thermal component into the total flux is less than 25\%. The resulting spectral index of the integrated
radio flux from the CMZ, defined as $\beta=-\Delta(\log S_\nu)/\Delta(\log\nu)$, is $\beta^{325~{\rm MHz}}_{1.4~{\rm GHz}} =
0.17 \pm 0.01$, $\beta^{1.4~{\rm GHz}}_{4.5~{\rm GHz}} = 0.58 \pm 0.01$, and $\beta^{4.5~{\rm GHz}}_{8.5~{\rm GHz}} = 1.14
\pm 0.01$. The measured spectrum $S_\nu$ is showed in Figure~\ref{fig1}. These measurements can be naturally interpreted as
a result of depletion of CR electrons in the CMZ, whose spectrum below the break is much harder than, e.g., in the Galactic
local medium \citep[see][]{biss19}. The solid line shows the synchrotron spectrum predicted by our theoretical model, as
described below in Section~\ref{expected}.

\begin{figure}[h]
\centering
\includegraphics[width=.95\columnwidth,clip=]{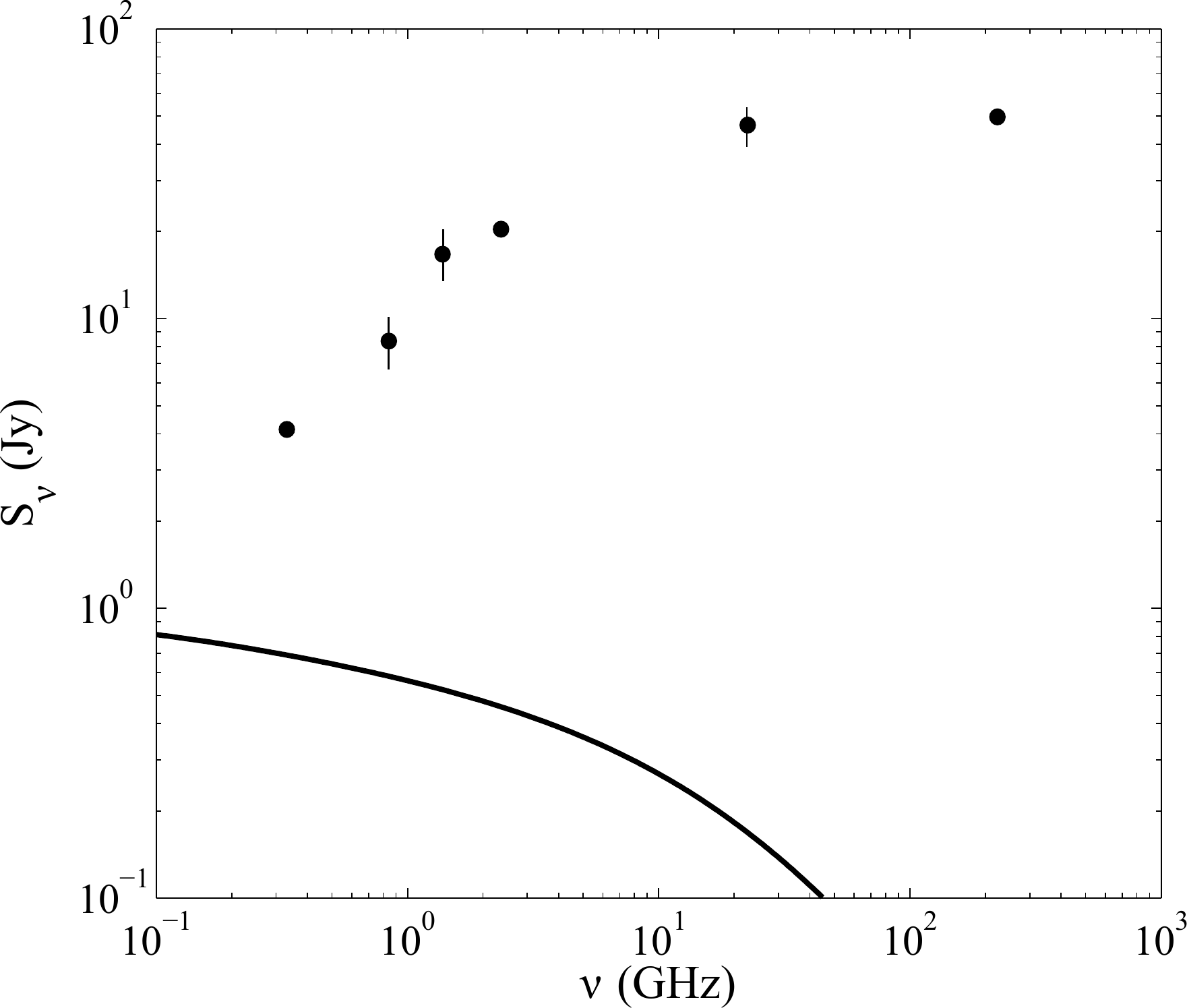}
\caption{The spectrum of radio emission reported by \citet{proth08} from the central region of Sgr B2
(bullets). One can see a $S_\nu\propto \nu^2$ scaling at lower frequencies,
indicating thermal nature of the emission. For comparison, we also plot the synchrotron emission spectrum (solid line)
calculated for the central region of Sgr B2 from our model (see Section~\ref{expected}).} \label{fig2}
\end{figure}

A stronger effect of CR depletion would be expected for very dense individual clouds. One of the suitable candidates is the
giant molecular cloud complex Sgr B2 \citep[e.g.,][]{hut95}. While its average column density is about $N_{{\rm H}_2}^{\rm
CMZ}$, in the central region of the Sgr B2 within $5~{\rm pc}\times2.5~{\rm pc}$ we have $N_{{\rm H}_2}^{\rm B2}= 10^{24}-
10^{25}$~cm$^{-2}$ \citep{schm16}, i.e., 1--2 orders of magnitude higher than the average CMZ value.

\citet{proth08} \citep[see also][]{jon11,meng19} found that the radio emission from the central region of Sgr B2 is
dominated by thermal emission of HII gas at frequencies above $\sim0.3$~GHz, as shown in Figure~\ref{fig2}. Thus, unlike the
CMZ, no indications of synchrotron-like emission from the Sgr B2 region were observed. We note that the same is true for
several other very dense molecular clouds, located outside of the Galactic center: the complexes G333.125--0.562 and G333
\citep[][]{jon08} and IRAS15596 \citep[][]{jon14}. On the other hand, we point out that the radio spectra from the Sgr B2
region reported in \citet{yus07,yus16} indicate a transition to non-thermal emission at the lowest observed frequencies of
$\sim0.3$~GHz.

Below we are going to find out the values of parameters required for the radio emission from modulated spectrum of electrons
to match the observations of the CMZ. We are also going to investigate the ratio between thermal and non-thermal emission
components for individual molecular clouds, to determine the feasibility to detect the non-thermal component.

\section{Spectrum of CR electrons inside dense clouds}
\label{spectrum}

\citet{ivlev} and \citet{dog3} developed the model of CR modulation in a diffuse envelope of a dense molecular cloud. It was
assumed that energy losses in the envelope can be neglected and the that the cloud absorbs a flux of CR protons\footnote{For
simplicity, here we only consider CR protons; heavier nuclei can be straightforwardly added, see \citet{dog3}.} with a
certain velocity $u_p$. For the proton spectrum $f_p(p)$ in the momentum space (normalized such that $\int f_p(p)dp$ is the
total number density of CRs), this gives the boundary condition at the cloud edge in the following form:
\begin{equation}
S_p(p) = u_pf_p^{\rm (c)}(p) \,, \label{eq:u_boundary}
\end{equation}
where $S_p(p)$ is the flux of CR protons into the cloud, $p$ is their momentum and $f_p^{\rm (c)}(p)$ is their spectrum at
the cloud edge. The flux velocity $u_p$ exceeds the Alfv\'{e}n velocity $v_{\rm A}$, which is a necessary condition of
excitation of MHD waves by CRs. In the Appendix we calculate $u_p$ for different propagation regimes in the cloud.

Given Equation~(\ref{eq:u_boundary}), the flux of CR protons entering the cloud was estimated as \citep{dog3}
\begin{equation}
S_p(p) = \frac{v_{\rm A}f_p^{(0)}(p)}{1-(1 - v_{\rm A}/u_p) e^{-\eta_p(p)}} \,, \label{eq:s_general}
\end{equation}
where $f_p^{(0)}(p)$ is the proton spectrum in the ISM, which is assumed to decrease with $p$ faster than $\propto p^{-1}$.
The ``diffusion depth'',
\begin{equation}\label{etha}
\eta_p(p) = \int \limits_0^{z_0} \frac{v_{\rm A}dz}{D_p(z,p)} \,,
\end{equation}
characterizes the relative importance of CR advection and diffusion in the envelope \citep{ivlev}: diffusion determines the
flux as long as $\eta_p\ll1$, whereas for $\eta_p\gtrsim1$ advection dominates and $S_p\approx v_{\rm A}f_p^{(0)}$. The
value of $\eta_p(p)$ depends on the proton diffusion coefficient $D_p(z,p)$, which is determined by the spectrum of
self-generated turbulence. The integration limit $z_0(p)$ is the (momentum-dependent) outer border of the diffusion zone in
the envelope \citep[see Figure~2 of][]{ivlev}.

If only a small fraction of penetrating protons is attenuated in a cloud, we can assume that their density in the cloud
(where the self-generated turbulence is absent) remains almost constant. Therefore, Equations~(\ref{eq:u_boundary}) and
(\ref{eq:s_general}) allow us to estimate an average spectrum of any CR species inside the cloud, provided $D(z,p)$ is
known.

\citet{ivlev} obtained the expression for $\eta_p(p)$ assuming the excitation-damping balance for MHD waves, which is valid
as long as $\eta_p\lesssim1$. It was shown \citep{dog3} that CR protons give the major contribution to the self-generated
turbulence, while the contribution of other CR species can be largely ignored. The growth rate of MHD waves is proportional
to the diffusion component of the flux \citep{ivlev}, and hence the excitation-damping balance can be written as
\begin{equation}
\frac{\pi^2e^2 v_{\rm A}}{m_pc^2\Omega} \:p\left[S_p(p)-v_{\rm A}f_p^{(0)}(p)\right] = \nu \,,
\label{balance}
\end{equation}
where $\Omega = eB/(m_pc)$ is the proton gyrofrequency scale, $B$ is the magnetic field strength, $e$ is the elemental
charge, $m_p$ is the mass of the proton, and $\nu$ is the damping rate of MHD waves due to ion-neutral collisions. Using
Equation~(\ref{eq:s_general}), this can be solved for $\eta_p$, which yields
\begin{equation}
\eta_p(p) = \ln \left[\left(1+ \frac{\pi^2e^2 v_{\rm A}^2}{ m_pc^2\Omega\nu}\:pf_p^{(0)}(p)\right)\delta_p\right] \,,
\label{eq:eta0_protons}
\end{equation}
where $\delta_p =1-v_{\rm A}/u_p>0$. Since $\eta_p\geq0$ by definition, the maximum momentum of modulated protons, $p_{\rm
ex}$, is determined from condition that the argument of the logarithm is equal to unity \citep[see][]{dog3}. Only protons
with $p< p_{\rm ex}$ are able to resonantly excite MHD waves in the envelope (which leads to their efficient scattering),
while protons with larger momenta propagate through the envelope freely, without scattering.

Substituting Equation~(\ref{eq:eta0_protons}) into Equation~(\ref{eq:s_general}) and assuming $\delta_p \approx 1$, one can
estimate the average proton spectrum inside a cloud,
\begin{equation}
u_pf^{\rm (c)}_p(p)\propto
 \left\{
\begin{array}{ll}
p^{-1}\,, & {\rm if}~ p_{\rm A} \leq p \leq p_{\rm ex}\,, \vspace{.2cm}\\
f_p^{(0)}(p)\,, & {\rm if}~p\geq p_{\rm ex}\,.
\end{array}\right.
\label{eq:Np_clouds}
\end{equation}
In the Appendix we demonstrate that $u_p$ does not practically depend on $p$. Therefore, the spectrum has a universal form
$f^{\rm (c)}_p(p)\propto p^{-1}$ for $p_{\rm A} \leq p \leq p_{\rm ex}$, with $p_{\rm A}$ estimated from the condition
$\eta_p(p_{\rm A})\sim1$. We note that the universal dependence $S_p(p)\propto p^{-1}$, occurring for $0<\eta_p\lesssim1$
(where $S_p$ is dominated by diffusion), follows directly from the excitation-damping balance, Equation~(\ref{balance}). For
$p\lesssim p_{\rm A}$, where $\eta_p$ exceeds unity and according to Equation~(\ref{eq:s_general}) the proton flux converges
to the advection component $v_{\rm A}f_p^{(0)}(p)$, we again obtain $f^{\rm (c)}_p(p)\propto f^{\rm (0)}_p(p)$.

Now we can estimate the electron spectrum inside the cloud. Indeed, the diffusion coefficient for both protons and electrons
is calculated as
\begin{equation}
D(p) \approx \frac{vB^2}{6\pi^2 k^2 W(k)} \,,
\end{equation}
where $v(p)$ is the particle velocity, $k(p) = eB/(pc)$ is the resonant wavenumber, and $kW(k)$ is the total energy density
of the proton-generated turbulence per unit $\log k$ range (i.e., $8\pi kW(k)/B^2$ is the dimensionless energy density).
Therefore, the electron flux $S_e(p)$ is described by Equation~(\ref{eq:s_general}) where the proton interstellar spectrum
and the flux velocity are replaced with the respective electron values, $f_e^{(0)}$ and $u_e$, while the electron diffusion
depth is
\begin{equation}
\eta_e(p) = \eta_p(p)\frac{v_p(p)}{v_e(p)} \,,
\label{eq:eta0_pe}
\end{equation}
as follows from Equation~(\ref{etha}). Assuming for simplicity that both electrons and protons are relativistic, $p \gg
m_pc$, and setting $\delta_e \approx 1$, we obtain the electron spectrum inside the cloud,
\begin{equation}
u_ef^{\rm (c)}_e(p)\propto
 \left\{
\begin{array}{ll}
{\displaystyle p^{-1}\frac{f_e^{(0)}(p)}{f_p^{(0)}(p)}\,,} & {\rm if}~p_{\rm A} \leq p \leq p_{\rm ex}\,, \vspace{.2cm}\\
f_e^{(0)}(p)\,, & {\rm if}~ p \geq p_{\rm ex}\,,
\end{array}\right.
\label{eq:Ne_clouds}
\end{equation}
with the same $p_{\rm A}$ and $p_{\rm ex}$ as in Equation~(\ref{eq:Np_clouds}). For relativistic CRs, their interstellar
spectra are well described by power-law dependencies, $f_{p,e}^{(0)}(p)\propto p^{-\gamma_{p,e}}$, with $\gamma_p\approx2.7$
and $\gamma_e\approx3.2$ \citep[see e.g.][]{biss19}. Thus, the electron spectrum inside a cloud scales as $f_e^{\rm
(c)}(p)\propto p^{-1+\gamma_p-\gamma_e}\approx p^{-1.5}$ (again, for $u_e=$~const) within the ``universal'' momentum range
of $p_{\rm A} \leq p \leq p_{\rm ex}$.

\section{Expected radio spectrum from dense clouds}
\label{expected}

Using the results of previous section as well as the expressions for $u_p$ and $u_e$ presented in the Appendix, we evaluate
the electron spectrum $f^{\rm (c)}_e(p)$ inside the cloud. This allows us to compute the total synchrotron emissivity
$\mathcal{P}_\nu$ of ultra-relativistic electrons per unit volume \citep{gs65},
\begin{equation}
\mathcal{P}_\nu\approx \frac{\sqrt{3}e^3B_\perp}{m_ec^2}
\int\limits dp\: f^{\rm (c)}_e(p)\frac{\nu}{\nu_*}\int\limits_{\nu/\nu_*}^{\infty}dx\:K_{5/3}(x)\,,
\label{Fnu}
\end{equation}
where $K_\alpha(x)$ is the McDonald function and
\begin{equation}
\nu_*(p)=\frac{3eB_\perp}{4\pi m_ec}\left(\frac{p}{m_ec}\right)^2 \,,
\end{equation}
is determined by the plane-of-the-sky magnetic field strength $B_\perp$. The emissivity scales as $\mathcal{P}_\nu\propto
B^{(\gamma+1)/2}\nu^{-(\gamma-1)/2}$ for a power-law electron spectrum $f^{\rm (c)}_e(p)\propto p^{-\gamma}$, with the
contribution of individual electrons peaked at a frequency of $\nu(p)\approx0.3\nu_*(p)$ \citep{gs65}. This indicates that
(i) the synchrotron emission is primarily generated in denser regions, where the magnetic field is stronger, and (ii) the
emission spectrum exhibits a break due to CR modulation. According to Equation~(\ref{eq:Ne_clouds}), at higher frequencies,
corresponding to unmodulated electrons with $p\gtrsim p_{\rm ex}$, the emission varies as $\mathcal{P}_\nu \propto
\nu^{-(\gamma_e-1)/2}\approx \nu^{-1.1}$, while at lower frequencies the spectrum becomes harder, $\mathcal{P}_\nu \propto
\nu^{-(\gamma_e-\gamma_p)/2}\approx \nu^{-0.25}$.

For a quantitative comparison with available observations from the CMZ \citep{yus13}, we assumed $n_{{\rm H}_2}=
50$~cm$^{-3}$ for the density of molecular hydrogen in the diffuse envelope, $B=0.1$~mG for the average magnetic field (we
note that this value does not practically affect the resulting radio spectrum), and used the solar abundance of carbon ions
to calculate the Alfv\'{e}n velocity and the ion-neutral damping in the envelope. The cloud column density, which determines
the value of $p_{\rm ex}$, was set to $N_{{\rm H}_2}^{\rm CMZ} = 10^{23}$~cm$^{-2}$. For the proton or electron spectra
outside the envelope we used the model form suggested in \citet{ivlev15},
\begin{equation}
f_{e,p}^{(0)}(p)={\rm const}\frac{E^\alpha}{(E_0+E)^\beta}\,,
\label{IS}
\end{equation}
where $E(p)$ is the proton or electron kinetic energy, $E_0=500$~MeV (same for both species), $\alpha=-0.8~(-1.5)$ and
$\beta=1.9~(1.7)$ for protons (electrons). These values of $E_0$, $\alpha$, and $\beta$ were selected in \citet{ivlev15} to
provide the upper bound for the available observational data on the CR ionization rate in diffuse gas \citep{ind12}.
Constant factors were chosen to ensure that the spectrum of relativistic protons is the same as in the local interstellar
medium \citep{acero16}, while the electron interstellar spectrum was enhanced by a factor of 10. The synchrotron emission
flux was calculated by assuming that the radiation mainly comes from dense regions occupying $\sim10$\% of the total CMZ
volume, $V_{\rm rad} \sim 0.1V_{\rm CMZ} \sim 10^{61}$~cm$^{3}$. The magnetic field in the dense regions was set 5 times
stronger than the average field, $B_{\rm dense} = 0.5$~mG. The total emission flux at a distance $R_{\rm CMZ}\approx 8$~kpc
is then obtained as $S_\nu=\mathcal{P}_\nu V_{\rm rad}/(4\pi R_{\rm CMZ}^2)$.

Furthermore, to explore the role of CR absorption in dense clouds, we considered a model case of the central region of Sgr
B2, corresponding to the observations reported by \citet{proth08}. In this case, we set the cloud column density to $N_{{\rm
H}_2}^{\rm B2} = 10^{24}$~cm$^{-2}$, keeping the other parameters unchanged.

\begin{figure}[h]
\centering
\includegraphics[width=\columnwidth,clip=]{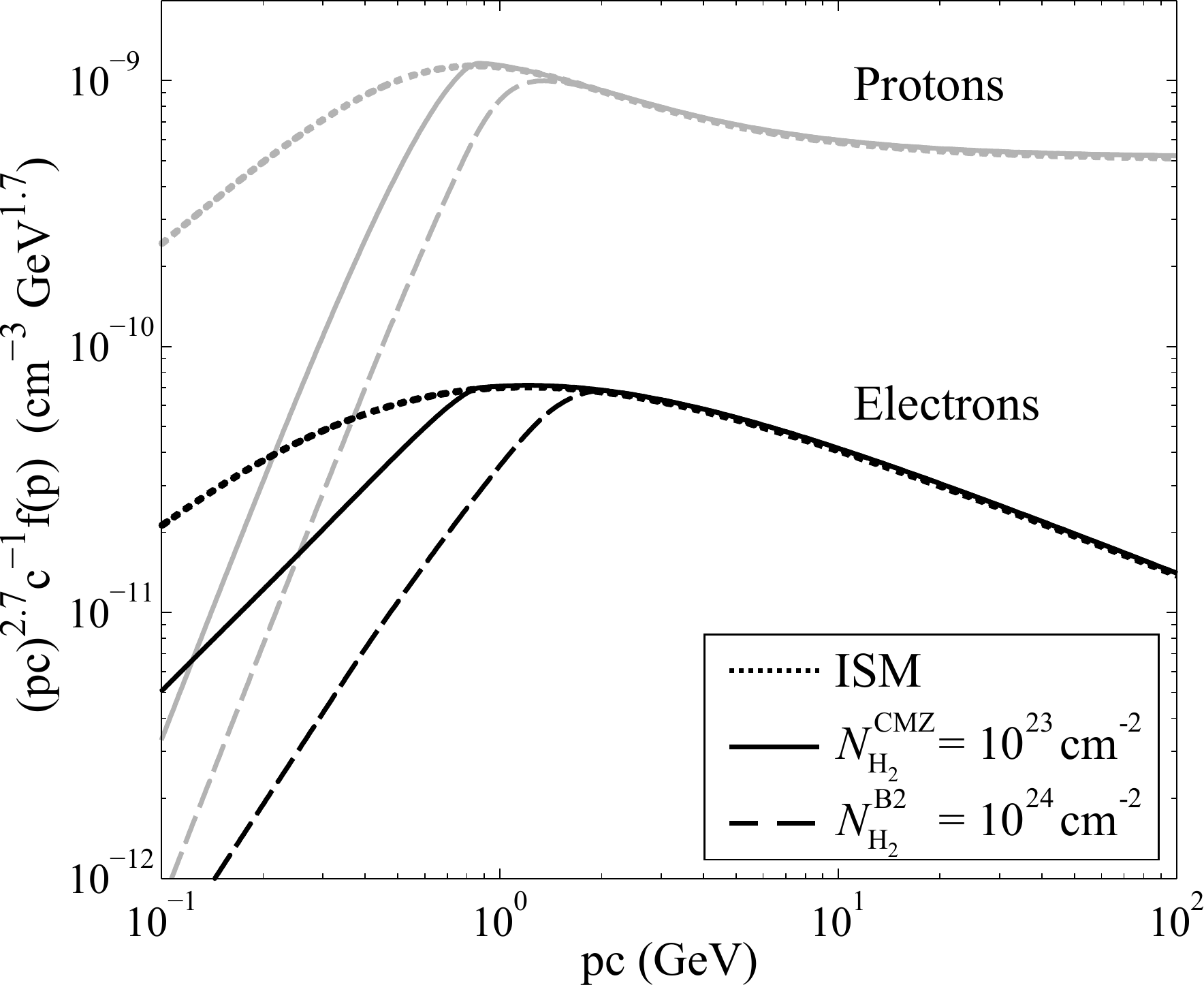}
\caption{Modulated spectra of CR electrons and protons inside CMZ clouds. The solid lines represents the spectra inside
an average CMZ cloud with the column density of $N_{{\rm H}_2}^{\rm CMZ} = 10^{23}$~cm$^{-2}$, the dashed lines
correspond to the central region of Sgr B2 with $N_{{\rm H}_2}^{\rm B2} =10^{24}$~cm$^{-2}$. For comparison, the
dotted lines show the interstellar (unmodulated) spectra. For the shown energy range, the densities of secondary electrons
(not displayed) are negligible compared to the respective modulated electron densities.} \label{fig3}
\end{figure}

Figure~\ref{fig3} shows the results of our calculations, displaying self-modulated spectra of CR protons (grey lines) and
the resulting modulated electron spectra (black lines). We see that the modulation for an average CMZ cloud (solid lines)
leads to a depletion of sub-GeV electrons, but the depletion magnitude is less than a factor of $\sim3$ for
$pc\gtrsim0.2$~GeV. For the above parameters, this modulated energy range approximately corresponds to the frequency range
between $\sim0.3$~GHz and $\sim3$~GHz, i.e., up to the spectral break seen in the radio data depicted by the bullets in
Figure~\ref{fig1}. Therefore, the solid line in Figure~\ref{fig1}, representing the calculated synchrotron flux, provides a
good agreement with the data, both in terms of the position of the spectral break and the slopes above and below the break.

The depletion of CR electrons in denser clouds, such as the central region of Sgr B2, starts at higher energies, as shown in
Figure~\ref{fig3} by the dashed line. The depletion magnitude in this case is up to a factor of $\sim20$ (for the same
energy range). The corresponding synchrotron flux, plotted by the solid line in Figure~\ref{fig2}, is too low to compete
with the thermal emission. Note that the secondary electrons were estimated to have a negligible ($\lesssim10\%$)
contribution to the above results.

We point out that the predicted modulation of CR electrons can also be reflected in the continuum gamma-ray emission. A
complete sky survey of the gamma-ray emission below 100 MeV was performed by COMPTEL imaging telescope \citep{shon93}. The
spectrum was analyzed by \citet{strong96}, who concluded that the bremsstrahlung radiation in this range must be primarily
produced by Galactic CR electrons. Therefore, the emission may indeed be affected by the electron depletion, but the spatial
and energy resolution of the COMPTEL data is too poor to observe the effect. The spectral break is expected at around
$0.1-0.3$~GeV, both for proton- and electron-generated emission, and thus next-generation gamma-ray telescopes operating in
this energy range, such as AMEGO \citep[][]{mcen20} and e-ASTROGAM \citep[][]{ang21}, may be able to observe the depletion
effect. The necessary condition for that would be their ability to resolve individual molecular clouds. In case of the CMZ,
the required spatial resolution is about $1^\circ$.

\section{Conclusions}
\label{conc}

We analyzed non-thermal radio emission from dense CMZ regions, and demonstrated that the observed spectral features can well
be interpreted in the framework of our recent theory of CR self-modulation in molecular clouds \citep{ivlev,dog3}. Our
conclusions are summarized as following:
\begin{itemize}
\item Self-modulation of CR protons with energies below a certain threshold $E_{\rm ex}$, determined by the column
    density of a cloud, generates a universal ($\propto E^{-1}$) flux of protons entering the cloud. This leads to a
    significant depletion of both the proton and electron densities below $E_{\rm ex}$. Proton-proton collisions are
    generally believed to be the main source of gamma-ray emission from clouds. However, due to low spatial and energy
    resolutions of Fermi-LAT at the relevant sub-GeV energies, the effect of proton depletion cannot be observed
    (although next-generation gamma-ray telescopes may potentially be able to resolve the emission). On the other hand,
    thanks to a high resolution of radio telescopes in the $\sim$~GHz range, it is possible to observe individual clouds
    and, thus, to detect the synchrotron emission of modulated electrons;
\item Assuming power-law spectra of relativistic interstellar protons and electrons, with the respective spectral
    indices $\gamma_p$ and $\gamma_e$, the electron (momentum) spectrum inside a cloud scales as $f_e(p)\propto
    p^{-1+\gamma_p-\gamma_e}$ for $p<p_{\rm ex}$ and $f_e(p)\propto p^{-\gamma_e}$ for $p>p_{\rm ex}$. This break
    naturally leads to a spectral break in the synchrotron emission;
\item The radio continuum from the CMZ is evidently non-thermal \citep[see][]{yus13} and, due to its large average
    column density of $N_{{\rm H}_2}^{\rm CMZ}\sim 10^{23}$~cm$^{-2}$, the CMZ is one of the most promising targets for
    studying the CR self-modulation. The value of $E_{\rm ex}(N_{{\rm H}_2}^{\rm CMZ})\sim 1$~GeV ensures the
    synchrotron spectral break at $\sim1$~GHz;
\item The observed radio spectrum from the CMZ can be approximated by a double power-law dependence with
    $S_\nu(\nu)\propto \nu^{-0.17}$ below $\sim1$~GHz and $S_\nu(\nu)\propto \nu^{-1.14}$ above $\sim1$~GHz (see
    Figure~\ref{fig1}). We demonstrate that this spectrum is well described by our theory of CR self-modulation,
    assuming the interstellar proton and electron spectra in the CMZ have the shape of the local CR spectra, with the
    electron density enhanced by a factor of 10;
\item A stronger depletion effect is expected for individual CMZ clouds, whose column densities are substantially larger
    than $N_{{\rm H}_2}^{\rm CMZ}$. For example, the expected depletion magnitude is a factor of $\sim5$ higher for the
    central region of Sgr B2 (see Figure~\ref{fig3}), where the column can exceed $N_{{\rm H}_2}^{\rm CMZ}$ by 1--2
    orders of magnitude \citep[see][]{proth08}. However, the radio emission from the Sgr B2 region is evidently
    dominated by the thermal component at all observed frequencies (see Figure~\ref{fig2}), and therefore new
    observations at lower frequencies are needed to verify the depletion effect.
\end{itemize}

We would like to thank an anonymous referee for constructive and stimulating suggestions. The authors are grateful to Farhad
Yusef-Zadeh for critical reading of the manuscript and useful comments, and to Roland Crocker for discussions. The work is
supported by Russian Science Foundation via the Project 20-12-00047.

\appendix
\section*{CR flux velocity at the edge in dense cloud}
\label{appx2}

To solve the problem of CR depletion self-consistently, one needs to evaluate the flux velocities $u_{p,e}$ of CR protons
and electrons entering dense clouds with energies in a range from several hundreds of MeV to dozens of GeV. For these
energies, the main mechanism of energy loss is bremsstrahlung for electrons, and pion production in nuclear collisions for
protons. Both process can well be treated as catastrophic, since CR particles lose a significant fraction of their energy in
each collision.

In the absence of scattering, CRs propagate along the magnetic field lines. Their spectrum is then described by the
following equation:
\begin{equation}
v\mu \frac{\partial f}{\partial z} + \frac{f}{\tau} = 0 \,,
\label{eq:appx_nosc_eq}
\end{equation}
with the boundary condition $f(0,p,\mu) = f_0(p,\mu)$. Here, $\mu$ is the pitch angle and $\tau^{-1} = v\sigma_{\rm loss}
n_{{\rm H}_2}$ is the collision rate with H$_2$ molecules, determined by the cross section $\sigma_{\rm loss}$ for the
relevant catastrophic loss mechanism (bremsstrahlung or pion production). If the CR spectrum at the cloud boundaries is
isotropic, we have $f_0(p)= f^{\rm (c)}(p)/2$. The resulting solution is
\begin{equation}
f(p,z,\mu) = f_0(p)\exp (-\sigma_{\rm loss} n_{{\rm H}_2}z/\mu) \,,
\end{equation}
and the flux velocity at the boundaries can be estimated as
\begin{equation}
u =\frac12\int \limits_0^1 v\mu\: d\mu - \frac12\int \limits_{0}^1v\mu\exp (-\sigma_{\rm loss} N_{{\rm H}_2}/\mu)\: d\mu \,,
\end{equation}
where $N_{{\rm H}_2}=n_{{\rm H}_2}L$ is the column density of a cloud of size $L$. For a typical case $\sigma_{\rm
loss}N_{{\rm H}_2} \ll 1$, after a linear expansion of the second term we obtain
\begin{equation}
u \approx \frac12v\sigma_{\rm loss} N_{{\rm H}_2}\,.
\label{eq:appx:unoscp_br}
\end{equation}

While CRs traveling inside dense clouds cannot normally trigger MHD waves because of a strong ion-neutral damping
\citep[see][]{kuls69}, the gas itself may be turbulent and thus the magnetic field lines can be strongly tangled.
Propagation of CRs is then described as an ``effective'' diffusion \citep[see e.g. the review of][]{hen12}. The energy of
magnetic field fluctuations with $\delta B\gg B_0$ is concentrated near the correlation length, $l_{\rm corr}$, and the
effective diffusion coefficient of relativistic CRs is estimated as $D\sim cl_{\rm corr}/3\sim10^{28}$~cm$^2$~s$^{-1}$,
assuming $l_{\rm corr}\approx 0.5$~pc for molecular clouds with the size of several pc \citep{dog15}. We stress that this
consideration does not require magnetic field to be perfectly frozen into the turbulent gas. In dense clouds, non-ideal MHD
effects are dominated by a combination of magnetic diffusion due to finite conductivity and ion-neutral friction
\citep{dog87,ist13}, but turbulent gas motions generate significantly tangled magnetic field lines also under such
conditions.

In a turbulent cloud with the effective diffusion coefficient $D$, the CR propagation is described by a simplified diffusion
equation,
\begin{equation}
D\frac{\partial^2 f}{\partial z^2} - \frac{f}{\tau} = 0 \,.
\label{DD}
\end{equation}
The boundary conditions are $\left.f\right|_{z = 0} = \left.f\right|_{z = L} = f_0$ and the solution is
\begin{equation}
f(p,z) = f_0(p)\frac{\exp\left(\frac{L-z}{\sqrt{D\tau}}\right) + \exp\left(\frac{z}{\sqrt{D\tau}}\right)}{1 +
\exp\left(\frac{L}{\sqrt{D\tau}}\right)} \,.
\end{equation}
The flux velocity is
\begin{equation}
u=\left.\frac{D}{f_0}\frac{\partial f}{\partial z}\right|_{z=L}=\sqrt{\frac{D}{\tau}}\tanh\left(\frac{L}{2\sqrt{D\tau}}\right)
\approx \frac12v\sigma_{\rm loss} N_{{\rm H}_2} \,,
\label{eq:appx:unoscp_br2}
\end{equation}
where the last (approximate) equality requires the condition $\sigma_{\rm loss}N_{{\rm H}_2}\ll l_{\rm corr}/L$ to be
satisfied. In the opposite limit, the velocity approaches the value of
\begin{equation}
u\approx v\sqrt{\frac13\frac{l_{\rm corr}}{L}\sigma_{\rm loss} N_{{\rm H}_2}}\,,
\end{equation}
where we substituted $D=vl_{\rm corr}/3$. We conclude that $u$ is not affected by the presence of turbulence in ``thin''
clouds (where the flux of CRs entering a cloud from one side is practically compensated by the opposite flux, arriving from
the other side). On the other hand, for sufficiently large $N_{{\rm H}_2}$ or/and small $l_{\rm corr}/L$, the flux velocity
is smaller if a cloud is turbulent (since the diffusive propagation naturally suppresses the entering flux, while the
compensation due to the opposite side is strongly attenuated). It is noteworthy that Equation~(\ref{eq:appx:unoscp_br2})
would also describe an (unlikely) situation if MHD fluctuations, leading to CR scattering, were present in a cloud: in this
case the CR propagation is characterized by a certain (momentum-dependant) diffusion coefficient, and hence the above
analysis remains applicable.

We use the bremsstrahlung cross section from \citet{blumenthal70} and the pion production cross section from \citet{ahar96}
to estimate the flux velocity of electrons and protons, respectively. In both cases, the condition $\sigma_{\rm loss}N_{{\rm
H}_2}\ll 1$ is well satisfied for column densities below $\sim 10^{25}$~cm$^{-2}$. Therefore, unless $l_{\rm corr}/L$ is too
small, Equations~(\ref{eq:appx:unoscp_br}) and (\ref{eq:appx:unoscp_br2}) yield the same expression for $u$, irrespective of
whether the cloud is quiescent or turbulent. Furthermore, since the cross sections only weakly depend on the energy of
relativistic CRs, we conclude that the flux velocities can be considered as constant.

\end{document}